\documentclass[journal]{IEEEtran}
\ifCLASSINFOpdf
\else
\fi

\usepackage[dvips]{color}
\usepackage{amsfonts, amsmath, mathrsfs, amsbsy, amssymb, dsfont, color}
\usepackage{graphicx}

\newcommand\prefixtext[1]{%
  \ifvmode\else\\\@empty\fi
  \noalign{%
    \penalty0%
    \vbox{\mathstrut}%
    \penalty10000%
    \vskip-\baselineskip
    \penalty10000%
    \vbox to 0pt{%
      \normalbaselines
      \ifdim\linewidth=\columnwidth
      \else
        \parshape\@ne
        \@totalleftmargin\linewidth
      \fi
      \vss
      \noindent#1\par}%
      \penalty10000%
      \vskip-\baselineskip}%
      \penalty10000}

\newcommand{\qed}{\nobreak \ifvmode \relax \else
      \ifdim\lastskip<1.5em \hskip-\lastskip
      \hskip1.5em plus0em minus0.5em \fi \nobreak
      \vrule height0.75em width0.5em depth0.25em\fi}

\DeclareMathAlphabet{\mathpzc}{OT1}{pzc}{m}{it}
\newcommand{\comment}[1]{}
\def\({\left(}
\def\){\left)}

\def\({\left(}
\def\){\right)}
\def\[{\left[}
\def\]{\right]}
\def\BEq{\begin{eqnarray}}
\def\EEq{\end{eqnarray}}
\def\BE*{\begin{eqnarray*}}
\def\EE*{\end{eqnarray*}}
\def\BA{\begin{array}}
\def\EA{\end{array}}

\def\Nn{\nonumber}
\def\0{\mathbf{0}}
\def\1{\mathbf{1}}

\def\and{\prefixtext{and}}

\usepackage{amsfonts}
\usepackage{amsmath,bm}
\usepackage{mathrsfs}
\usepackage{subfigure}
\usepackage{graphicx}
\usepackage{amssymb}
\usepackage{stfloats}
\usepackage{arydshln}
\usepackage{multirow}
\usepackage{booktabs}
\usepackage{algorithm}
\usepackage{algorithmic}
\usepackage[colorlinks, citecolor=red]{hyperref}

\begin{document}
%
\title{A Survey on Matrix Completion: Perspective of Signal Processing}
%
%
%

\author{Xiao~Peng~Li,
        Lei~Huang,~\IEEEmembership{Senior Member,~IEEE,}
        Hing~Cheung~So,~\IEEEmembership{Fellow,~IEEE,}
        and Bo~Zhao,~\IEEEmembership{Member,~IEEE}
        }

\maketitle

\begin{abstract}
    Matrix completion (MC) is a promising technique which is able to recover an intact matrix with low-rank property from undersampled/incomplete data. Its application varies from wireless communications, traffic sensing to integrated radar and communications, and thereby receives much attention in the past several years. There are plenty of works addressing the behaviors and applications of MC methodologies. This work provides a comprehensive review for MC approaches from the perspective of signal processing. In particular, the MC problem is first grouped into seven optimization problems in the light of different occasions and formulations to help readers understand MC algorithms comprehensively. Next, five representative types of optimization algorithms solving the MC problem are reviewed. Furthermore, simulation results demonstrate the empirical performance of different types of MC optimization problem. Ultimately, five different application fields of MC, including two potential applications, are described and evaluated.

\end{abstract}


%
\IEEEpeerreviewmaketitle

\section{Introduction}
    \IEEEPARstart{D}{uring} the past few years, matrix completion (MC) has received increasing interest worldwide for its unique property and numerous applications in traffic sensing~\cite{1, 2}, integrated radar and communications~\cite{3}, image inpainting~\cite{4}, system identification~\cite{5}, multi-task learning~\cite{6,7} and so on. Subsequent to compressed sensing, MC is another significant technology utilizing sparse property to process data. Sparsity, in CS, means that the signal of interest contains lots of zero elements in a specific domain. However, in MC, it indicates that the singular value vector of the original matrix is sparse. In other words, the matrix is low-rank.

    MC is able to restore the original signal $\bm X$ from a fragmentary signal $\bm X_\Omega$ (or called the undersampled/incomplete signal), where $\Omega$ is a subset containing 2D coordinates of sampled entries. The undersampled signal $\bm X_\Omega$ can be expressed as
    \begin{equation}
    \label{E}
        \bm X_\Omega = \bm H_\Omega\odot\bm X + \bm N
    \end{equation}
    where all of variables belong to $\mathbb{R}^{m\times n}$, $\odot$ is the element-wise multiplication operator, $\bm H_\Omega$ and $\bm N$ are the sampling matrix and noise matrix, respectively. Note that $\bm H_\Omega$ is a binary matrix, which are drawn from a random uniform distribution to ensure at least one 1-element in each row and column~\cite{8}. Furthermore, it is assumed that the original signal $\bm X$ has the low-rank or approximately low-rank property~\cite{8}.

    Low-rank property of signals is ubiquitous in real-world applications. For instance, the received signal, in MIMO radars system, is of low-rank. This is because the targets and clutters in the cell under test (CUT) are sparse in space domain. The number of targets and clutters in the echoes corresponds to the rank of original signals, which is usually much less than the numbers of transmit and receive antennas. Another example is the image data matrix. The main information conveyed by the data matrix is dominated by some largest singular values, whereas the remaining smallest singular values can be taken as zero without losing major information. Thus, the image data matrix has an approximately low-rank structure.

    As shown in Fig 1, three panels stand for the distribution of singular values of the original image, the original image and the low-rank image, respectively. The matrix of original image owns 349, but most of them are almost equal to zero, as can be observed in the left panel of Fig. 1. In other words the largest ten singular values are enough to represent the original image.
    \begin{figure}[!htbp]
    \begin{minipage}[b]{0.5\linewidth}
      {\includegraphics[scale=0.35]{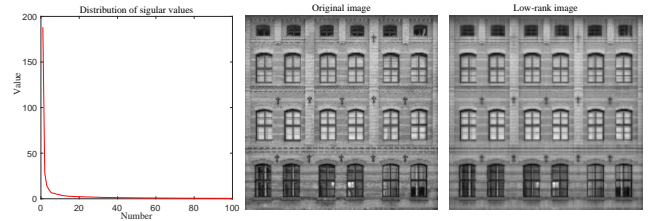}}
    \end{minipage}
    \caption{ Utilizing 10 largest singular values to represent the original image.}
    \end{figure}

    In the pioneering work of Cand${\rm \grave{e}}$s and Recht~\cite{8}, it is proposed to utilize rank minimization problem to restore the original signal $\bm X$. The MC problem under the noise-free environment is formulated as
    \begin{equation}
        \label{E0}
        \min\limits_{\bm{M}} {\rm rank}(\bm{M}),\ {\rm s.t.}\ \bm{M}_\Omega = \bm{X}_\Omega
    \end{equation}
    where $\bm M \in \mathbb{R}^{m\times n}$, $\bm X_\Omega = \bm H_\Omega\odot\bm X$, and $\bm{M}_\Omega$ denotes the projection on $\Omega$. When the sampled signal is corrupted by noise, there is a need to constrain the noise level within an appropriate range. As a result, the MC problem can be expressed as
    \begin{equation}
        \label{E26}
        \min\limits_{\bm{M}} {\rm rank}(\bm{M}),\ {\rm s.t.}\ \left \|\bm{M}_\Omega - \bm{X}_\Omega\right \|_F\leq\delta
    \end{equation}
    where $\bm X_\Omega$ is defined by~(\ref{E}), $\|\cdot\|_F$ denotes the Frobenius norm of a matrix and $\delta > 0 $ is a tolerance parameter that limits the fitting error.  Unfortunately, the rank minimization problem is NP-hard since all algorithms for exactly solving~(\ref{E0}) and (\ref{E26}) are doubly exponential in dimension ${\rm max}⁡(m,n)$ in both theory and practice. This is why all state-of-the-art algorithms attempt to solve the approximate problem of rank minimization.
    \begin{table*}[t]\label{tableI}
    \begin{center}
    \renewcommand\arraystretch{1.25}
    \caption{a summary of matrix completion methods}
    \vspace{0.5em}\centering
    \begin{tabular}{p{1.7cm}|p{1.6cm}|p{0.8cm}|p{2.2cm}|p{1.8cm}|p{2.5cm}|p{2.8cm}}
    \hline
        \multicolumn{7}{c}{Rank minimization}
        \\\hline\hline
        \multicolumn{5}{c|}{Normal situation} & \multicolumn{2}{c}{Outlier situation} \\
        \hline
        \multicolumn{3}{c|}{Nuclear norm minimization}   &\multirow{2}*{Matrix factorization}  &\multirow{2}*{\shortstack{Minimum rank \\approximation}}  &\multirow{2}*{\centering $\ell_p$-norm minimization} &\multirow{2}*{Adaptive outlier pursuing}\\\cline{1-3}
        Semidefinite programming  &\multirow{1}{1.9cm}[-2pt]{Nuclear norm relaxation}    &Robust PCA   &~ &~&~\\
        \hline
    \end{tabular}
    \end{center}
    \end{table*}
    Fazel~\cite{55} proved that nuclear norm is the convex envelope of rank, which turns out to be a convex relaxation, and in turn enables one to efficiently solve the issue of rank minimization in MC. This convex relaxation is akin to the relaxation of $\ell_0$ minimization to $\ell_1$ minimization in CS~\cite{56}. Subsequently, Cand${\rm \grave{e}}$s and Recht~\cite{8} proposed to solve the rank minimization problem (\ref{E0}) by the nuclear norm minimization problem, given as
    \begin{equation}
    \label{E1}
    \min\limits_{\bm{M}} \left \|\bm{M}\right \|_*,\ {\rm s.t.}\ \bm{M}_\Omega = \bm{X}_\Omega
    \end{equation}
    where $\|\cdot\|_*$ is the nuclear norm of a matrix. More significantly, Cand${\rm \grave{e}}$s and Tao~\cite{57} have theoretically proved that the original signal $\bm{X}$ with the strong incoherence property can be perfectly recovered with high probability via solving the problem in~(\ref{E1}).

    This article attempts to give an overview of MC methodologies, including different optimization problems and optimization algorithms, with emphasis on their principles as well as differences. Meanwhile, we provide several examples to showcase the MC applications.

\section{MC Formulations}
    Various MC methodologies have been developed from different perspectives, with pros and cons. To facilitate readers, we present a brief summary of several well-known MC algorithms in Table I.
\subsection{Nuclear Norm Minimization}
\subsubsection{Semidefinite programming}
    The nuclear norm minimization problem (\ref{E1}) can be recast as a semidefinite programming problem~\cite{8}, resulting in
    \begin{align}
            &\min\limits_{\bm{M},\bm{W_1},\bm{W_2}} {\rm tr}(\bm{W_1}) + {\rm tr}(\bm{W_2})\Nn\\
            &{\rm s.t.}\ \bm{M}_\Omega = \bm{X}_\Omega,\ {\begin{bmatrix} \bm{W_1} & \bm{M}\\ \bm{M}^T & \bm{W_2} \end{bmatrix}}\succeq 0
    \end{align}\\
    where $\bm{W_1} \in \mathbb{R}^{m\times m}$ and $\bm{W_2} \in \mathbb{R}^{n\times n}$ are positive semidefinite, ${\rm tr}(\bm {W_1})=\sum_{i=1}^{m}(w_1)_{i,i}$, $\bm M^T$ is the transposed matrix of $\bm M$, $\bm X\succeq 0$ means $\bm X$ being positive semidefinite.

    There are several efficient algorithms to solve this semidefinite programming problem, including SDPT3~\cite{9} and SeDeMi~\cite{10}. However, these approaches are usually based on interior-point technique and their computational complexity can be as high as $O(p(m+n)^3  + p^2 (m+n)^2+p^3)$ flops, where $p$ denotes the number of known entries in $\bm{X}_\Omega$. Usually, they can only solve the $m\times n$ matrix with $m$ and $n$ being not larger than 100 on a moderate personal computer. For instance, put $p=0.3mn$, when $m=n=100$, the running time is about 1 minute; when $m=n=120$, the running time is approximately 5 minutes; while when $m=n\geq200$, the MATLAB will overflow. Readers could obtain more details about interior-point methods for nuclear norm approximation in~\cite{11}.
\subsubsection{Nuclear norm relaxation}
    Based on nuclear norm minimization problem, the singular value thresholding (SVT) approach proposed to use a proximal objective of nuclear norm minimization~\cite{12}, given as
    \begin{equation}
    \label{E2}
    \min\limits_{\bm{M}} \tau\left \|\bm{M}\right \|_*+\frac{1}{2}\left \|\bm{M}\right \|_F^2,\ {\rm s.t.}\ \bm{M}_\Omega = \bm{X}_\Omega
    \end{equation}
    where $\tau\geq0$. It is proved in~\cite{12} that minimizing (\ref{E2}) is analogous to minimizing (\ref{E1}) in the limit of large $\tau$. Note that the parameter $\tau$ provides a tradeoff between the nuclear norm and Frobenius norm. As $\tau$ becomes large, the optimization issue in (\ref{E2}) amounts to that in (\ref{E1}). In the end, the result of (\ref{E2}) can be obtained via solving its Lagrangian
    \begin{equation}
        \label{E4}
        L(\bm{M},\bm{Y})=\tau\left \|\bm{M}\right \|_*+\frac {1}{2}\left \|\bm{M}\right \|_F^2+\langle \bm{Y},\bm{M}_\Omega-\bm{X}_\Omega \rangle.
    \end{equation}

    To solve (\ref{E4}), Cai, $et$ $al$.~\cite{12} introduced a proximity operator associated with the nuclear norm. In particular, a soft-thresholding operator $D_\tau$ is introduced, which is defined as
    \begin{align}
        \label{E3}
        &D_\tau(\bm{Y}):=\bm{U}D_\tau(\bm S)\bm V^T\Nn\\
        &D_\tau(\bm S)={\rm diag}(\{(\sigma_i-\tau)_+\}_{1\leq i\leq r})
    \end{align}
    where $r$ is the rank of $\bm Y$, $\bm{Y}=\bm{U}\bm S\bm{V}^T$ is the singular value decomposition (SVD) of $\bm{Y}$ with $\bm S={\rm diag}(\{\sigma_i\}_{1\leq i\leq r})$, $\bm U \in \mathbb{R}^{m\times r}$ and $\bm V\in \mathbb{R}^{n\times r}$ being orthonormal matrices, and $t_+={\rm max}⁡(0,t)$. Here, it should be emphasized that many popular algorithms have utilized this operator to solve the nuclear norm minimization problem in the literature, say~\cite{13,14,15}, to name a few.

    \begin{table}[!hp]\label{tableII}
    \begin{center}
    \renewcommand\arraystretch{1.25}
    \caption{Algorithm for different ranks}
    \vspace{0.5em}\centering
    \begin{tabular}{p{3cm}p{3cm}}
    \hline
        $r$ &Algorithm\\\hline \hline
        $r\leq0.1n$ &PROPACK\\
        $0.1n< r\leq0.25n$ &Fast SVT\\
        $0.25n< r$ &Full SVD\\\hline
    \end{tabular}
    \end{center}
    \end{table}
    Notably, each iteration in solving (\ref{E4}) requires to calculate the SVD of $\bm Y$ and then obtain $D_\tau(\bm{Y})$. When the rank of $D_\tau(\bm{Y})$ is much lower than its dimension, partial SVD algorithms such as package PROPACK are extremely efficient. However, the partial SVD algorithm becomes less and less efficient as the rank of $D_\tau(\bm{Y})$ increases. To handle this problem, a fast SVT (FSVT) approach~\cite{13} is devised to directly compute $D_\tau(\bm{Y})$, avoiding SVD at each iteration. As a guideline, a summary of approaches to compute $D_\tau(\bm{Y})$ for different ranks is provided in Table~II.

    In order to handle the noisy situation, fixed point continuation with approximate SVD (FPCA)~\cite{14} and accelerated proximal gradient with line-search-like acceleration (APGL)~\cite{15} take noise into consideration, and relax the standard nuclear norm minimization problem into the least absolute shrinkage and selection operator (LASSO), given by:
    \begin{equation}
        \label{E5}
        \min\limits_{\bm{M}} \lambda\left \|\bm{M}\right \|_*+\frac{1}{2}\left \|\bm{M}_\Omega-\bm{X}_\Omega\right \|_F^2
    \end{equation}
    where $\lambda>0$ is the regularization parameter used to tradeoff the nuclear norm and Frobenius norm that corresponds to the power of residual between $\bm M_\Omega$ and $\bm X_\Omega$. Note that both FPCA and APGL utilize the soft-thresholding operator. In addition, FPCA uses the fixed point continuation and Bregman iterative methods to solve~(\ref{E5}), whereas APGL employs the accelerated proximal gradient approach to solve~(\ref{E5}) and incorporated the line-search strategy to accelerate the convergence.

    The aforementioned algorithms are constructed based on the standard nuclear norm, which try to  minimize all singular values simultaneously. Nevertheless, it is not appropriate to minimize all singular values in some scenarios. For instance, the largest singular values of image matrix describe the major edge and texture information, and thus need to be maintained during the nuclear-norm minimization. To cope with this problem, a truncated nuclear norm regularization (TNNR) approach has been proposed in~\cite{16} to improve the accuracy of MC. Only the smallest ${\rm min}(m,n)-r$ singular values are minimized in the TNNR method, which is formulated as
    \begin{equation}
        \label{E6}
        \min\limits_{\bm{M}}\left \|\bm{M}\right \|_r,\ {\rm s.t.}\ \bm{M}_\Omega = \bm{X}_\Omega
    \end{equation}
    where $||\bm{M}||_r= \sum_{i=r+1}^{{\rm min}⁡(m,n)}{\sigma_i (\bm M)}=||\bm M||_*-{\rm tr}(\bm{AMB}^T)$, $\bm A=[\bm u_1,...,\bm u_r[^T$ and $\bm B=[\bm v_1,...,\bm v_r[^T$. Here, $\bm u_r$ and $\bm v_r$ are the left and right singular vectors of $\bm{M}$, respectively. Because alternating direction method of multipliers (ADMM) owns decomposability of dual ascent with the superior convergence properties of the method of multipliers, TNNR employs ADMM to solve (\ref{E6}). The TNNR-ADMM scheme reformulates (\ref{E6}) as
    \begin{align}
        \min\limits_{\bm M, \bm W}\left \|\bm{M}\right \|_*-{\rm tr}(\bm{AMB}^T)\Nn\\
        {\rm s.t.}\ \bm M = \bm W,\ \bm W_\Omega = \bm X_\Omega.
    \end{align}

    Note that TNNR-ADMM has the constraint of $\bm W_\Omega = \bm X_\Omega$ which means that the sampled entries with noise in $\bm X_\Omega$ will be intactly retained in $\bm M$, so it is less effective for noisy data. To circumvent this problem, TNNR-APGL algorithm was suggested in~\cite{16} by utilizing the accelerated proximal gradient line search (APGL) technique. The TNNR-APGL technique relaxes~(\ref{E6}) as
    \begin{equation}
        \label{E22}
        \min\limits_{\bm M, \bm W}\left \|\bm{M}\right \|_*-{\rm tr}(\bm{AMB}^T)+\frac{\lambda}{2}\left \|\bm M_\Omega-\bm X_\Omega\right \|_F^2
    \end{equation}
    where $\lambda>0$. TNNR-APGL is suitable for noisy environment on account of the third term in (\ref{E22}).

    Although TNNR algorithm is able to significantly enhance the recovery performance in the MC problem, it requires to determine the rank of matrix in advance.
\subsubsection{Robust PCA}
    Lin, $et$ $al$.~\cite{20} considered the MC problem as a special case of robust principal component analysis (PCA) problem and formulated it as
    \begin{equation}
        \label{E7}
        \min\limits_{\bm M}\left \|\bm M\right \|_*,\ {\rm s.t.}\ \bm M+\bm S = \bm{X}_\Omega,\ \bm S_\Omega = 0
    \end{equation}
    where $\bm S\in \mathbb R^{m\times n}$ is a sparse matrix. The inexact augmented Lagrange multipliers (IALM)~\cite{20} solves the augmented Lagrange version of (\ref{E7}) to obtain the result $\bm M$. However, the approach in (\ref{E7}) does not consider the noisy environment due to $\bm S_\Omega=0$, thereby prohibiting its applications.

    To improve the accuracy of MC, weighted nuclear norm minimization (WNNM)~\cite{21} introduced different weights to singular values to avoid to shrink all singular values equally. WNNM is more flexible than TNNR since the larger the weight is, the smaller the singular value becomes. Under the critical situation, WNNM can also maintain the largest singular values corresponding to zero weights. The weighted nuclear norm of a matrix $\bm M$ is defined as
    \begin{equation}
        \left \|\bm M\right \|_{w,*}=\sum_i w_i\sigma_i(\bm M)
    \end{equation}
    where $\bm w=[w_1,...,w_n ]^T$ with $w_i\geq 0$ being a non-negative weight assigned to $\sigma_i(\bm M)$. Based on the weighted nuclear norm, the variant of robust PCA for MC was devised in~\cite{21}, which is formulated as
    \begin{equation}
        \label{E8}
        \min\limits_{\bm M}\left \|\bm M\right \|_{w,*},\ {\rm s.t.}\ \bm M+\bm S=\bm X_\Omega,\ \bm S_\Omega = 0
    \end{equation}

    It should be noticed that although the standard robust PCA for low-rank matrix recovery is able to process impulsive noise, the robust PCA for MC in~(\ref{E7}) and~(\ref{E8}) is not robust against impulsive noise. The standard robust PCA is formulated as
    \begin{equation}
        \label{E23}
        \min\limits_{\bm L,\bm S} \left \|\bm L\right \|_*+\lambda\left \|\bm S\right \|_1,\ {\rm s.t.}\ \bm L+\bm S = \bm D
    \end{equation}
    where $\bm L\in \mathbb R^{m\times n}$ is the target matrix with low-rank property, $\lambda>0$. Interestingly, $\bm S$ in the constraint of~(\ref{E23}) can be taken as impulsive noise added to $\bm L$. Accordingly, its sparse property can be characterized by the $\ell_1-$norm. Therefore, the standard robust PCA is robust against impulsive noise whereas its variant for tackling the MC problem does not retain this robustness. Actually, if the sampled entries in~(\ref{E7}) are corrupted by additive noise, the noise term cannot be suppressed due to $\bm S_\Omega = 0$. This is why the robust PCA for MC has a bad performance in the case of noise, not to mention impulsive noise.
\subsection{Minimum Rank Approximation}
    The forementioned methodologies for solving the MC problems are devised based on the assumptions of noiseless or noisy samples. As a matter of fact, we cannot foreknow that whether the data are corrupted by noise or not. To cope with this problem, atomic decomposition for minimum rank approximation (ADMiRA)~\cite{30} proposed to solve the MC problem via alternative formulation of rank minimization problem, called minimum rank approximation problem, which is
    \begin{equation}
        \label{E12}
         \min\limits_{\bm M}\left \| (\bm M)_\Omega-\bm X_\Omega\right \|_F^2,\ {\rm s.t.}\ {\rm rank}(\bm{M})\leq r
    \end{equation}
    where $r$ is the bound of rank. The advantage of this optimization problem is that it considers noiseless and noisy cases. It is also more suitable for the situation where the original matrix is not of exactly low-rank but can be approximately of low rank.

    ADMiRA is developed in the framework of orthogonal matching pursuit and it, in each iteration, first searches for $2r$ components and then obtains a $r$-rank matrix by carrying out SVD. As a result, it exhibits low computational efficiency for large-dimensional matrices. To cope with this problem, singular value projection (SVP)~\cite{31} proposed to employ the singular values projection method to solve (\ref{E12}). At the same time, it also utilizes a Newton-type step to improve the accuracy and convergence. Besides, the variants of ADMiRA have been put forward in~\cite{32,33} to tackle the minimum rank approximation problem.
\subsection{Matrix Factorization}
    Although the MC approaches are capable of offering superior performance by tailoring the nuclear norm minimization criterion, they suffer from low computational efficiency and limited scalability in big-scale data. To circumvent this issue, the matrix factorization (MF)~\cite{22} was proposed to solve the MC problem without SVD. The basic idea behind the MF methodology is to utilize two low-rank matrices to represent objective matrix with an assumption that the rank of original matrix is known. The low-rank matrix fitting (LMaFit)~\cite{22} is the first algorithm employing the MF technique to solve the MC problem. Mathematically, the problem is
    \begin{equation}
        \label{E9}
        \min\limits_{\bm U, \bm V, \bm Z}\left \|\bm U\bm V^T-\bm Z\right \|_F^2,\ {\rm s.t.}\ \bm Z_\Omega=\bm X_\Omega
    \end{equation}
    where $\bm U \in\mathbb{R}^{m\times r}$, $\bm V \in\mathbb{R}^{n\times r}$, and $\bm Z \in\mathbb{R}^{m\times n}$ with $r$ being a predicted rank of the objective matrix. Then, it employs a successive over-relaxation technique to solve the Lagrange version of (\ref{E9}). Despite LMaFit is able to obtain an accurate solution, it cannot be globally  optimal due to its non-convex function. Alternating minimization for matrix completion (AltMinComplete)~\cite{23} is a variant of LMaFit, in which the optimization problem becomes
    \begin{equation}
        \label{E10}
        \min\limits_{\bm U, \bm V}\left \|(\bm U\bm V^T)_\Omega-\bm X_\Omega\right \|_F^2
    \end{equation}
    To boost the convergence of the optimization procedure, the block coordinate descent method (also called alternative minimizing method) has been tailored in~\cite{23} to solve (\ref{E10}). It is the first work which theoretically investigates the global optimality on the MF-based MC approach.

    In order to further enhance the performance of MF-based MC approach, OptSpace~\cite{24} factorizes the objective matrix as $\bm M=\bm U\bm S \bm V^T$, and solves the following optimization problem on the Grassmann manifold
    \begin{equation}
        \label{E11}
        \min\limits_{\bm {U}, \bm {V}}\min\limits_{\bm S}\left \|(\bm U \bm S \bm V^T)_\Omega-\bm X_\Omega\right \|_F^2
    \end{equation}
    where $\bm U \in\mathbb{R}^{m\times r}$ and $\bm V \in\mathbb{R}^{n\times r}$ satisfy $\bm U^T\bm U=m\bm I$ and $\bm V^T \bm V=n\bm I$. Moreover, $\bm S \in \mathbb{R}^{r\times r}$ is a diagonal matrix. To obtain a smooth objective function, OptSpace needs to simultaneously search the row and column spaces which, however, cannot guarantee a globally optimal solution as there may exist barriers along the search path. To fix this problem, subspace evolution and transfer (SET)~\cite{25} factorizes matrix into two low-rank matrices in the form of $\bm M=\bm U\bm V$, yielding the following optimization problem
    \begin{equation}
        \min\limits_{\bm {U}}\min\limits_{\bm V}\left \|(\bm U \bm V^T)_\Omega-\bm X_\Omega\right \|_F^2
    \end{equation}
    where $\bm U\in \mathbb{R}^{m\times r}$ is the orthonormal matrix, and $\bm V\in \mathbb{R}^{n\times r}$ with $r$ being much less than $\min(m,n)$. Compared with OptSpace, SET only searches for a column (or row) space. Furthermore, to guarantee the result being a globally optimal solution, SET employs a mechanism to detect barriers and transfers the estimated column (or row) space from one side of the barrier to another. Subsequently, various variants of the MF-based MC approach have been addressed in~\cite{26,27,28,29}.
\subsection{\texorpdfstring{$\ell_p$}~-Norm Minimization}
    It should be pointed out that the Euclidean distance metric $\|\cdot\|_2^2$ ($\|\cdot\|_F^2$ or trace for matrix case) is able to accurately describe the variance of independent and identically distributed (IID) Gaussian noise. However, for impulsive noise which usually corrupts the received data in real-world applications, the $\ell_2$-norm cannot exactly characterize the behaviors of both impulsive and Gaussian noises. It is easy to understand this statement because the $\ell_2$-norm may seriously amplify the power of impulsive noise, which is much larger than the power of Gaussian noise. This thereby motivates one to exploit other metrics for the impulsive noise scenario. For a matrix $\bm R$, $\ell_p$-norm is defined as
    \begin{equation}
        \| \bm R \|_p=\left( \sum_{i,j}\left | [\bm R]_{i,j}\right |^p \right)^\frac{1}{p}
    \end{equation}
    where $[\bm R]_{i,j}$ is the element of $\bm R$.

    It is well known that $\ell_p$-norm with $0<p<2$ is able to resist outlier, and thereby has been widely adopted to handle the impulsive noise. However, few articles explain why it can resist impulsive noise. Here, we provide an explaination to help readers comprehend this property. Consider a minimization problem
    \begin{equation}
        \label{E21}
        \min\|\bm R\|_p^p,\ {\rm s.t.}\ \bm R=\bm M- \bm X, 0< p\leq 2
    \end{equation}
    where $\bm R$ is the residual matrix between $\bm M$ and $\bm X$, $\|\bm R\|_p^p =\sum_{i,j} |[\bm R]_{i,j}|^p$. Notice that $|[\bm R]_{i,j}|^p$ is the residual penalty term, and their sum stands for the total penalty. Different choices of $\bm M$ lead to different residuals, and eventually various approaches can be devised.

    Roughly speaking, $|[\bm R]_{i,j}|^p$ measures the level of our dislikes of $[\bm R]_{i,j}$. If $|[\bm R]_{i,j}|^p$ is very small, it does not affect the recovery performance. If $|[\bm R]_{i,j}|^p$ becomes large, however, it is indicated that we have to handle strong dislikes for these large residuals. Dislikes correspond to the values we need to minimize. For instance, compared with $|[\bm R]_{i,j}|$, $|[\bm R]_{i,j}|^2$ magnifies residuals, especially the residuals associated with outlier. In other words, to minimize the total residual, $|\cdot|^p$-norm~$(0<p<2)$ pays more attention to minimize large residuals, i.e., outlier. Consequently, $\ell_p$-norm ($p=1$) has a better performance than $\ell_2$-norm.

    The $\ell _p$-regression ($\ell _p$-reg) algorithm~\cite{34} combines the MF technique and $\ell _p$-norm to solve the MC problem, which is formulated as
    \begin{equation}
        \label{E13}
        \min\limits_{\bm U \bm V}\left \| (\bm U \bm V^T)_\Omega-\bm X_\Omega\right \|_p^p,\ {\rm s.t.}\ 0< p\leq 2.
    \end{equation}

    To tackle the distributed frame and big data efficiently, it utilizes the alternating minimization strategy was suggested in~\cite{34} to solve~(\ref{E13}).

    As a variant of the $\ell_p$-norm based alternating minimization, the alternating projection (AP) algorithm was put forward in~\cite{35}. Unlike the standard alternating minimization scheme, the AP approach formulates MC problem as a feasibility problem. More specifically, it firstly defines the following two sets
    \begin{align}
        \label{E14}
        S_r&:=\left \{\bm M|{\rm rank}(\bm M)\leq r\right\}\\
        \label{E15}
        S_p&:=\left \{\bm M|\left \|\bm M_\Omega -\bm X_\Omega \right \|_p^p\leq \varepsilon_p\right\}
    \end{align}
    where (\ref{E14}) and (\ref{E15}) are the low-rank set and fidelity constraint set, respectively. The constant $r$ is the estimated rank of $\bm M$ and $\varepsilon_p>0$ is a small tolerance parameter determined by the noisy matrix. Then, the AP algorithm finds the resultant $\bm M$ located in the intersection of $S_r$ and $S_p$ via the alternating projection method.

    It should be pointed out that, although the AP and $\ell _p$-reg algorithms are able to provide superior recovery performance, they both required to know the rank of matrix $\bm M$, which might not be available in real-world implementations. Besides, the noise parameter $\varepsilon_p$ in the AP algorithm is calculated from the noise-only matrix, which, however, incurs more overhead in a practical system.
\subsection{Adaptive Outlier Pursuing}
    Adaptive outlier pursuing (AOP) method is also able to resist outlier since it can detect the position of the outlier. Yan $et$ $al$.~\cite{58} proposed an algorithm which utilizes AOP technique to solve the MC problem under impulsive noise. This algorithm is called Riemannian trust-region for MC with AOP (RTRMC-AOP) and as
    \begin{align}
        \label{E28}
        \min\limits_{\bm U, \bm V, \bm\wedge} &\left\|\bm{\Lambda}\odot((\bm{UV}^T)_\Omega-\bm X_\Omega)\right \|_F^2\Nn\\
        &{\rm s.t.} \left \| \bm I - \bm\Lambda \right \|_1\leq K
    \end{align}
    where $\bm U \in\mathbb{R}^{m\times r}$ satisfying $\bm U^T\bm U=m\bm I$, $\bm V \in\mathbb{R}^{n\times r}$, $K$ being the number of outliers in $\bm X_\Omega$, and $\bm{\Lambda} \in\mathbb{R}^{m\times n}$ being a binary matrix. Then, RTRMC-AOP employs alternating minimization method to solve (\ref{E28}), which splits $\bm U, \bm V, \bm\Lambda$ in two groups, exactly $\bm\Lambda$ and $\bm U$, $\bm V$, and then minimizes these two groups of parameters alternately. The emphasis of RTRMC-AOP is to update $\bm\Lambda$ which can be calculated by
    \begin{align}
        \label{E29}
        \bm \Lambda_{i,j}=\left\{
        \begin{aligned}
        1 & , {\rm if} (i,j)\in \Omega, ((\bm U\bm V^T)_{i,j}-M_{i,j})^2\leq \tau, \\
        0 & , {\rm otherwise}.\\
        \end{aligned}
        \right.
    \end{align}
    where $\tau$ is the $K$th largest term in the set of $\{(\bm U\bm V^T)_{i,j}-M_{i,j})^2,(i,j)\in \Omega\}$. In practice, the value of $K$ is unknown. To solve this problem, Yan, $et$ $al$. also proposed a technique to update $K$ during calculating $\bm U, \bm V, \bm\wedge$ such that it is able to recover the exact matrix in the case of unknown $K$ with high probability.
\section{Algorithms}
    Numerous algorithms can be employed to solve the MC problems. In this section, we will review five main types of optimization approaches which vary from gradient to non-gradient schemes. These optimization methods include gradient descent (GD), accelerated proximal gradient (APG), Bregman iteration (BI), block coordinate descent (BCD) and alternating direction method of multipliers (ADMM). Table~III provides a summary of them.
    \begin{table}[!ht]\label{tableIV}
    \begin{center}
    \renewcommand\arraystretch{1.25}
    \caption{a summary of optimization methods}
    \vspace{0.5em}\centering
    \begin{tabular}{p{1cm}p{1cm}p{1cm}|p{1cm}p{1cm}}
    \hline
        \multicolumn{5}{c}{Optimization algorithms} \\\hline \hline
        \multicolumn{3}{c|}{Gradient} &\multicolumn{2}{c}{Non-gradient}   \\\hline
        GD   &APG     &BI   &BCD   &ADMM\\\hline
    \end{tabular}
    \end{center}
    \end{table}
\subsection{Gradient}
\subsubsection{Gradient descent}
    GD is the most fundamental optimization method for unconstrainted minimization problem. Consider an unconstrainted minimization problem
    \begin{equation}
        \min\limits_{\bm X\in\mathbb{R}^{m\times n}} F(\bm X)
    \end{equation}
    where $F(\bm X)$ is a convex and smooth function and its gradient is $\nabla F(\bm X)$. Then the GD approach is described in Algorithm~1.
    \begin{algorithm}[htb]
    \caption{GD}
    \begin{algorithmic}[1]
    \REQUIRE Maximum iteration $N$, $\bm X^0$\\
    \FOR{ $k=0, 1,…, N$}
    \STATE $\bm X^{k+1}= \bm X^{k}-\delta \nabla F(\bm X^{k})$
    \ENDFOR
    \ENSURE $\bm X^{k+1}$\\
    \end{algorithmic}
    \end{algorithm}\\
    where $\delta$ is a step size. Usually, it is hard to select the appropriate step size $\delta$. If $\delta$ is sufficiently small such that the convergence can be guaranteed,
    but the speed of convergence turns out to be very slow. On the contrary, if $\delta$ is small, the convergence cannot be ensured.
\subsubsection{Accelerated proximal gradient}
    If $F(x)$ contains a non-smooth term, its gradient cannot be computed, leading to the inapplicability of GD-like approach. To bypass this problem, a proximal gradient (PG) algorithm was suggested in~\cite{41}. Subsequently, its convergence was boosted in~\cite{42} via Nesterov technique, ending up with the APG method. To be precise, the optimization problem in the APG algorithm
    is formulated as
    \begin{equation}
        \label{E27}
        \min\limits_{\bm X\in\mathbb{R}^{m\times n}} F(\bm X)=\min\limits_{\bm X\in\mathbb{R}^{m\times n}} J(\bm X)+H(\bm X)
    \end{equation}
    where $J(\bm X)$ is a convex and smooth function, whereas $H(\bm X)$ is a convex but non-smooth function. Before going deep into the APG algorithm, let us first briefly review the
    proximal operator. For $H(\bm X)$, the proximal operator is
    \begin{align}
        \label{E24}
        &{\rm prox}_{\delta H}(\bm X^k)=\min\limits_{\bm X^{k+1}}(H(\bm X^{k+1})+\frac{1}{2\delta}\left \|\bm X^{k+1}- \bm X^k\right \|_F^2)
    \end{align}
    where $\delta >0$ compromises between minimizing $H(\cdot)$ and being near to $\bm X^k$. The proximal operator is obtain a $\bm X^{k+1}$ which satisfies $H(\bm X^{k+1})<H(\bm X^k)$ and is akin to GD method. After a finite number of iterations, we can get the $\bm X$ which minimizes the value of $H(\cdot)$.~\cite{41} proposed a PG method to solve~(\ref{E27}). Mathematically, the PG is expressed as
    \begin{align}
        \bm X^{k+1}:={\rm prox}_{\delta H}(\bm X^k-\delta  \nabla J(\bm X^k))
    \end{align}
    where PG first obtains the value $\dot{\bm X}^k$ such that $J(\dot{\bm X}^k)\leq J(\bm X^k)$ via $\bm X^k-\delta  \nabla J(\bm X^k)$ being the GD expression of $J(\cdot)$. Then combine the proximal operator to achieve $H(\bm X^{k+1})< H(\dot{\bm X}^k)$ so $F(\bm X^{k+1})\leq F(\bm X^k)$. Based on the PA, APG algorithm was devised in~\cite{42}, which is summarized in Algorithm 2.
    \begin{algorithm}[htb]
    \caption{APG}
    \begin{algorithmic}[1]
    \REQUIRE Maximum iteration $N$, $\bm X^0$, $\bm Y^0$ and $t_0=1$\\
    \FOR{ $k=0, 1,…, N$}
    \STATE $\bm X^{k+1}= {\rm prox}_{\delta H}(\bm Y^k-\delta \nabla J(\bm X^k))$
    \STATE $t_0=(1+\sqrt {1+4t_k^2})/2$
    \STATE $\bm Y^{k+1}= \bm X^{k}+\frac{t_k-1}{t_{t+1}}(\bm X^{k+1}-\bm X^k)$
    \ENDFOR
    \ENSURE $\bm X^{k+1}$\\
    \end{algorithmic}
    \end{algorithm}

    It should be pointed out that the accelerated variant of PG approach is not successive descent, and its convergence is thereby akin to the shape of ripples.
\subsubsection{Bregman iteration}
    As another type of approach to handle the non-smooth minimization, BI~\cite{37} is proposed to solve the equality-constrained minimization problem. Since Osher, $et$ $al$.~\cite{38} employed BI to address the total variation based image restoration problem, it has been widely extended to CS~\cite{39} and image deblurring~\cite{40}.
    It now becomes an efficient tool in solving the MC problem and can be utilized to tackle the general equality-constrained minimization problem, namely
    \begin{equation}
        \min\limits_{\bm X}H(\bm X),\ {\rm s.t.}\ C(\bm X)=0
    \end{equation}
    where $\bm X \in\mathbb{R}^{m\times n}$ and this equality constrained minimization problem can be translated into unconstrained minimization problem by relaxing the constraint, as follows
    \begin{equation}
        \label{E16}
        \min\limits_{\bm X}H(\bm X)+J(\bm X)
    \end{equation}
    where $J(\bm X)$ is smooth and convex, while $H(\bm X)$ is only convex. Before employing BI algorithm to solve (\ref{E16}), we share a concept of Bregman distance. For the convex function $H(\cdot)$, it is defined
    \begin{equation}
        D_H^P\left (\bm X, \bm Y \right )= H(\bm X)-H(\bm Y)-\langle \bm P,\bm X-\bm Y \rangle
    \end{equation}
    where $\bm P\in \partial H(\bm Y)$ is some sub-gradient in the sub-differential of $H(\cdot)$ at $\bm Y$. The main idea behind the BI approach is to construct the so-called Bregman distance in order to bypass non-differential point of $H(\cdot)$. In particular, one tries to find a set of sub-gradient of $H(\cdot)$ at $\bm Y$, such that the following Bregman distance can be minimized. The BI for solving (\ref{E16}) is summarized in Algorithm 3.
    \begin{algorithm}[htb]
    \caption{BI}
    \begin{algorithmic}[1]
    \REQUIRE Maximum iteration $N$, $\bm X^0=0$ and $\bm P^0$=0\\ 
    \FOR{ $k=0, 1,…, N$}
    \STATE $\bm X^{k+1}= {\rm arg}\min\limits_{\bm X\in\mathbb{R}^{m\times n}} D_H^{P^k}(\bm X, \bm X^k)+J(\bm X)$
    \STATE $\bm P^{k+1}=\bm P^k-\nabla J(\bm X^{k+1})$
    \STATE $k=k+1$
    \ENDFOR
    \ENSURE $\bm X^{k+1}$\\ 
    \end{algorithmic}
    \end{algorithm}

    Compared with GD and AGD strategies, BI algorithm has a faster convergence speed. Furthermore, GD requires to shrink the step size during iteration, while BI does not change the step size, avoiding the instability in parameter adjustment.
\subsection{Non-gradient}
    The forementioned three types of optimization methods are constructed by explicitly or implicitly computing the gradient of cost function. In some real-world implementations, however, it might be impossible to find the (approximate) gradient of objective function. This thereby motives one to find the non-gradient type of minimization strategy.
    \begin{table*}[t]\label{tableV}
    \begin{center}
    \renewcommand\arraystretch{1.25}
    \caption{comparison of different optimization algorithms}
    \vspace{0.5em}\centering
    \begin{tabular}{p{2.0cm}p{6.0cm}p{5.1cm}}
    \hline
        Algorithm  &Advantages  &Disadvantages\\\hline\hline
        \multirow{1}{1.5cm}[-1pt]{GD/APG}   &High efficiency of low-dimension matrix completion and fast speed of convergence.   &Require to compute SVD in matrix completion problem.\\\hline
        BI   &Fast speed of convergence, do not required to solve the exact solution of sub-problem. &Not suitable for distributed manner, low efficiency in big-scale problem.\\\hline
        \multirow{1}{1.5cm}[-2pt]{BCD} &Wide application, easy to operate, basic algorithm of matrix completion.   &Cannot ensure convergence in the case of non-smooth objective function.\\\hline
        \multirow{1}{1.5cm}[-5pt]{ADMM}   &Combine the merits between dual ascent and the method of multipliers, suitable for distributed form.  &\multirow{1}{5.1cm}[-5pt]{Low efficiency in the case of high accuracy.}\\\hline
    \end{tabular}
    \end{center}
    \end{table*}
\subsubsection{Block coordinate descent}
    \begin{algorithm}[htb]
    \caption{BCD}
    \begin{algorithmic}[1]
    \REQUIRE Maximum iteration $N$, $\bm X^0$,$\bm Y^0$ and $\bm Z^0$\\
    \FOR{ $k=0, 1,…, N$}
    \STATE $\bm X_0^{k+1}= {\rm arg}\min\limits_{\bm X^k}F(\bm X^k,\bm Y^k,\bm Z^k)$
    \STATE $\bm X^{k+1}=\omega \bm X_0^{k+1}+(1-\omega)\bm X^k$
    \STATE $\bm Y_0^{k+1}= {\rm arg}\min\limits_{\bm Y^k}F(\bm X^{k+1},\bm Y^k,\bm Z^k)$
    \STATE $\bm Y^{k+1}=\omega \bm Y_0^{k+1}+(1-\omega)\bm Y^k$
    \STATE $\bm Z_0^{k+1}= {\rm arg}\min\limits_{\bm Z^k}F(\bm X^{k+1},\bm Y^{k+1},\bm Z^k)$
    \STATE $\bm Z^{k+1}=\omega \bm Z_0^{k+1}+(1-\omega)\bm Z^k$
    \ENDFOR
    \ENSURE $\bm X^{k+1}, \bm Y^{k+1},\bm Z^{k+1}$\\ 
    \end{algorithmic}
    \end{algorithm}

    As the non-gradient type of minimization scheme, BCD~\cite{43} has been widely adopted to deal with large-scale optimization problem since it finds the optimal estimates of the parameters in a distributed manner, significantly enhancing the computational efficiency. The main principle behind the BCD algorithm is to optimize one parameter set while keeping other parameter sets unchanged at one time. For instance, given an unconstrained optimization problem
    \begin{equation}
        \min\limits_{\bm X,\bm Y,\bm Z} F(\bm X,\bm Y,\bm Z)
    \end{equation}
    one attempts to minimize $F(\bm X,\bm Y,\bm Z)$ firstly with respect to $\bm X$, while considering $\bm Y$ and $\bm Z$ to be known. The same procedure is then applied to $\bm Y$ and $\bm Z$. The BCD method is summarized in Algorithm 4.

    In algorithm 4,  $\omega \geq 1$ is a factor that is able to speed up convergence. When $\omega=1$, it is the standard BCD algorithm and solves steps 2, 4 and 6 alternately in Algorithm 4 to obtain $\bm X$, $\bm Y$ and $\bm Z$ directly. While $\omega>1$, it turns out to be the accelerated BCD, called nonlinear successive over-relaxation (SOR) algorithm. The parameter $\omega$ is able to tradeoff the new and legacy values in steps 2, 5 and 7 such that a more suitable value can speed up convergence of the objective function.

    The convergence behavior of the SOR algorithm for solving the MC problem has been studied in~\cite{22}. To ensure the convergence of BCD, it is required that $F(\bm X,\bm Y,\bm Z)$ is smooth. In addition, each component of $F(\bm X,\bm Y,\bm Z)$ is strong convex and Lipschitz continuous. If objective function is non-differentiable, however the convergence cannot be ensured.
\subsubsection{Alternative Direction Method of Multiplier}
    Note that BCD is devised to deal with non-constrained large-scale optimization issue. For constrained large-scale optimization problem, Gabay and Mercier~\cite{45} firstly introduced ADMM to tackle it. It is revealed that ADMM is very efficient in big-scale~\cite{43} and distributed~\cite{44} problems. According to the principle of ADMM, the constrained problem to be optimized can be expressed as
    \begin{align}
    \label{E19}
        &\min\limits_{\bm X,\bm Z} F(\bm X)+G(\bm Z)\Nn\\
        &{\rm s.t.}\ \bm {AX}+\bm{BZ}=\bm C
    \end{align}
    where $F(\bm X)$ and $G(\bm Z)$ are convex, $\bm X\in\mathbb{R}^{m\times r}$, $\bm Z\in\mathbb{R}^{n\times r}$, $\bm A \in\mathbb{R}^{p\times m}$, $\bm B \in\mathbb{R}^{p\times n}$ and $\bm C \in\mathbb{R}^{p\times r}$. The ADMM firstly converts (\ref{E19}) to the augmented Lagrangian
    \begin{align}
        L_\delta(\bm X,\bm Y,\bm Z)=&F(\bm X)+G(\bm Z)+\langle \bm Y^T,\bm {AX}+\bm{BZ}-\bm C\rangle\Nn\\
        &+\frac{\delta}{2}\left \|\bm {AX}+\bm{BZ}-\bm C\right \|_F^2
    \end{align}
    where $\delta>0$. Then, the BCD approach is employed to optimize $\bm X$, $\bm Y$ and $\bm Z$ separately. Algorithm 5 summarizes the ADMM approach.
    \begin{algorithm}[htb]
    \caption{ADMM}
    \begin{algorithmic}[1]
    \REQUIRE Maximum iteration $N$, $\bm X^0$,$\bm Z^0$ and $\delta$\\ 
    \FOR{ $k=0, 1,…, N$}
    \STATE $\bm X^{k+1}= {\rm arg}\min\limits_{\bm X^k}L_\delta(\bm X^k,\bm Y^k,\bm Z^k)$
    \STATE $\bm Z^{k+1}= {\rm arg}\min\limits_{\bm Z^k}L_\delta(\bm X^{k+1},\bm Y^k,\bm Z^k)$
    \STATE $\bm Y^{k+1}= Y^{k}+\delta(\bm {AX^{k+1}}+\bm{BZ^{k+1}}-\bm C)$
    \ENDFOR
    \ENSURE $\bm X^{k+1}, \bm Z^{k+1}$\\ 
    \end{algorithmic}
    \end{algorithm}

    Notice that ADMM combines the decomposability of dual ascent with the superior convergence property of the method of multiplier. On the other hand, $\bm X$ and $\bm Z$ are updated in an alternating fashion which accounts for the term of alternating direction. To fit big-scale computation and machine learning,~\cite{46} develops the asynchronous distributed ADMM whereas~\cite{47} derives the fast stochastic ADMM. Inspired by adaptive penalty strategy, Liu, $et$ $al$.~\cite{48} proposed a parallel splitting version of ADMM which can solve the multi-variable separable convex problem efficiently. Besides, it has been proved in~\cite{48} that the convergence of parallel ADMM can be guaranteed.

    The five types of algorithms above are the representative approaches to solve the MC problem. And their pros and cons are summarized in Table IV.
\section{Simulation Results}
    All simulations in this section are conducted on a personal computer with i7-6700, 3.4GHz and 8GB memory. The data to used are a synthetic matrix $\bm X\in\mathbb {R}^{150\times300}$ generated by the product of $\bm X_1\in \mathbb{R}^{150\times10}$ and $\bm X_2\in\mathbb{R}^{10\times300}$. All entries of these two matrices satisfy the standard Gaussian distribution with zero mean and unity variance. Meanwhile 45\% of the entries are selected from the matrix $\bm X$ randomly as the training matrix $\bm X_\Omega$. We evaluate six MC algorithms, including SVT, TNNR, IALM, OptSpace, SVP, and $\ell _p$-reg. And their codes are available online at https://github.com/hellofrankxp/Codes-of-MC.git. These MC methods cover all problems and optimization algorithms that can help readers better understand different problems and optimization algorithms. Performance is measured by the normalized root mean square error (RMSE), defined as
    \begin{equation}
    {\rm RMSE}(\hat{\bm M})=\sqrt{E\left\{\frac{\left\|\hat{\bm M}-\bm X\right\|_F^2}{\left\|\bm X\right\|_F^2} \right\}}
    \end{equation}
    where $\hat{\bm M}$ is the recovered matrix computed by a MC approach, and calculated based on 200 independent trials.
    \begin{figure}[!htbp]
    \begin{minipage}[b]{0.5\linewidth}
      {\includegraphics[width=9cm,height=6.777cm]{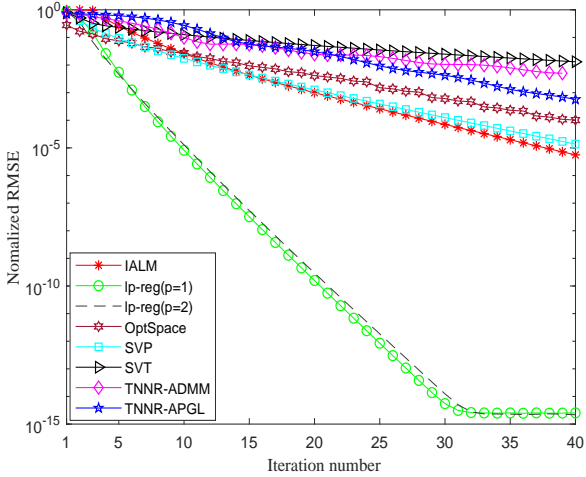}}
    \end{minipage}
    \caption{Normalized RMSE versus iteration number in noise-free case.}
    \end{figure}

    Fig. 2 shows the normalized RMSE versus iteration number in noise-free case. It is observed that $\ell _p$-reg with $p=1$ and $p=2$ have the best performance in term of accuracy and convergence. TNNR-ADMM and TNNR-APGL are better than SVT since they employ the TNNR technique to improve the accuracy. OptSpace and $\ell _p$-reg belong to the matrix factorization approach that does not relax rank function. As a result, they are superior to SVT, TNNR-ADMM and TNNR-APGL which belong to the nuclear norm relaxing problem. IALM and SVP have a moderate accuracy among the investigated algorithms.
    \begin{table*}[t]\label{tableVI}
    \begin{center}
    \renewcommand\arraystretch{1.25}
    \caption{cpu times for different matrix completion algorithms}
    \vspace{0.5em}\centering
    \begin{tabular}{ccccccccc}
    \hline
        Algorithm   &IALM   &SVP    &TNNR-APGL  &TNNR-ADMM  &$\ell_2$-reg   &SVT    &OptSpace   &$\ell_1$-reg\\\hline \hline
        Time(s) &0.3462 &0.3717 &0.6035 &1.5017 &1.7506 &3.1704 &3.2369 &9.0605\\\hline
    \end{tabular}
    \end{center}
    \end{table*}

    Let us now evaluate the MC algorithms for impulsive noise. Gaussian mixture model (GMM) has been widely used to simulate impulsive noise, and its PDF is defined as
    \begin{equation}
        p_v(v)=\sum_{i=1}^2{\frac{c_i}{\sqrt{2\pi}\sigma_i}{\rm exp}{(-\frac{v^2}{2\sigma_i^2}})}
    \end{equation}
    where $c_i\in[0,1]$ with $c_1+c_2=1$ is the probability and $\sigma_i^2$ is variance of the $i$th term. The total variance is $\sigma_v^2=c_1 \sigma_2^2+c_2 \sigma_2^2$. We set $\sigma_2^2\gg \sigma_1^2$ and $c_2<c_1$ which means that the large noise samples with bigger variance $\sigma_2^2$ and smaller probability $c_2$ can been considered as outliers mixed in Gaussian background noise with small variance $\sigma_1^2$. Thus, GMM can well model the impulsive noise with both outlier and Gaussian noise. Here, we set $\sigma_2^2=100\sigma_1^2$ and $c_2=0.1$, meaning that there are 10\% samples of outliers. Define the signal-to-noise ratio (SNR) as
    \begin{equation}
        {\rm SNR}=\frac{\left \|\bm X_\Omega\right \|_F^2}{|\Omega|\sigma_v^2}
    \end{equation}
    \begin{figure}[!htbp]
    \begin{minipage}[b]{0.5\linewidth}
      {\includegraphics[width=9cm,height=6.777cm]{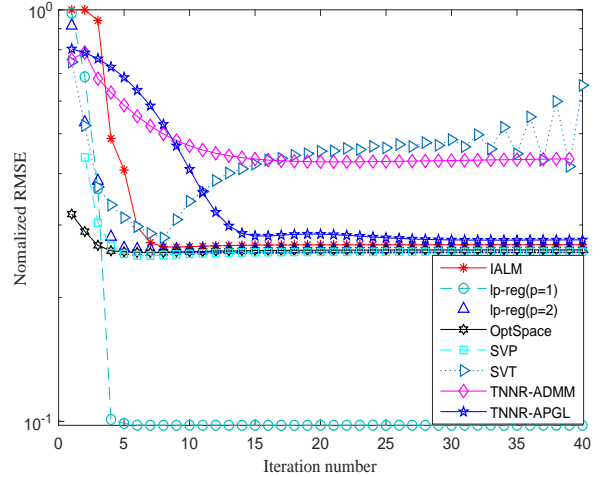}}
    \end{minipage}
    \caption{Normalized RMSE versus iteration number for GMM noise at 6dB case.}
    \end{figure}

    Fig. 3 plots the normalized RMSE against iteration number in the GMM noise case with SNR=6dB. $\ell _p$-reg with $p=1$ yields the highest accuracy and the fastest convergence , which indicates that the $\ell _1$-norm has a good performance of outlier resistance. while, SVT and TNNR-ADMM cannot stably converge to a good solution since they do not consider the noise in their problems. IALM, $\ell _2$-reg, OptSpace, SVP, and TNNR-APGL provide the moderate accuracy.

    Table V shows the CPU times of different algorithms when RMSE$<10^{-6}$ in the noiseless case. It can be observed that the CPU times of IALM, SVP and TNNR-APGL are less than one second. The CPU times of TNNR-ADMM and $\ell _2$-reg are 1.5017s and 1.7506s, respectively. SVT and OptSpace require a little bit more time, namely, around 3.2s. The $\ell _1$-reg consumes the most computational time though it is capable of offering superior recovery performance. Nevertheless, it could  be boosted by adopting the ADMM technique and its recovery performance might be maintained.
\section{Applications}
\subsection{SAR Imaging}
    Synthetic aperture radar (SAR) owns several advantages such as all-weather condition, high resolution, and interference suppression and so on. It has been widely utilized in military and civilian fields. However, the demand for high resolution inevitably increases the difficulty in transmission and storage of the raw data due to the data exploding. Yang $et$ $al$.~\cite{36} proposed to employ the MC technique to handle these two problems in the SAR system. Assume that $\bm X \in \mathbb R^{m\times n}$ is the receive data (raw data) in the SAR radar. After sub-sampling, the sparse data $\bm H\odot\bm X$ can be stored in the disk or transmitted to the base station. For instance, the base station receives the data $\bm Y=\bm H\odot\bm X+\bm N$ where $\bm N$ is the noise acquired during transmission, and then the raw data can be recovered via
    \begin{equation}
    \label{E25}
    \min\limits_{\bm M}\left \|\bm M\right \|_*,\ {\rm s.t.}\ \left \| \bm Y - \bm M_\Omega\right \|_F\leq \delta
    \end{equation}
    where $\delta>0$ is a tolerance parameter that controls the fitting error. After $\bm M$ obtained, it can be utilized to image instead of the raw data.
    \begin{figure}[!htbp]
    \begin{minipage}[b]{0.5\linewidth}
      {\includegraphics[scale=0.43]{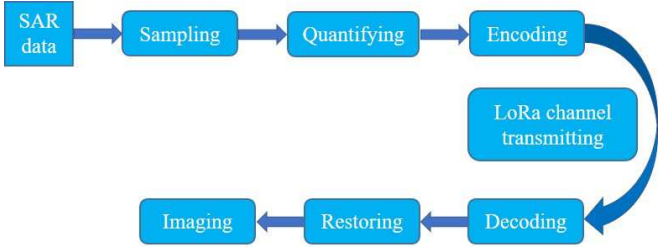}}
    \end{minipage}
    \caption{A flowchart of SAR data compression and recovery with MC technique. }
    \end{figure}

    Fig. 4 describes an experiment via employing MC technique to compress SAR data. Raw data is generated from the original image that is from the Sandia National Laboratories.
    Firstly, the raw data is sampled randomly and uniformly with 50$\%$ sampling rate. Then it is quantified and encoded by 16bits and Huffman method respectively. The wireless communication mechanism is LoRa which is widely adopted by enterprises because of its free. Ultimately, the data recovered by MC method is used to image instead of raw data.

    \begin{figure}[!htbp]
    \begin{minipage}[b]{0.5\linewidth}
      {\includegraphics[scale=0.4]{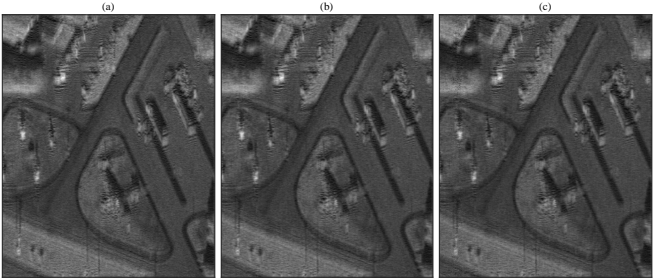}}
    \end{minipage}
    \caption{Performance of the SAR imaging by real data.}
    \end{figure}

    In Fig. 5, (a) is the original image that is from the Sandia National Laboratories. The remaining two are imaged from the recovered data restored in the cases of free noise and 10dB Gaussian noise respectively. As can be seen from (b), the data recovered from MC under noise-free conditions can be well imaged. In the meantime,  transmission efficiency can be improved by 42.37$\%$ since the transmission time of raw data is about 2845$s$, while the data after compression only costs 1640$s$. The performance of (c) indicates the MC technique is able to filter Gaussian noise and generate the clear data.

    \begin{figure}[!htbp]
    \centering
    \begin{minipage}[b]{0.5\linewidth}
      {\includegraphics[scale=0.4]{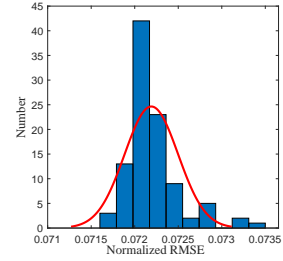}}
    \end{minipage}
    \caption{Distribution of normalized RMSE based on 100 independent trials.}
    \end{figure}

    Fig. 6 shows effect of different sampling matrices on normalized RMSE in noise-free case. Its mean is 0.0722 and standard deviation is $3.075\times 10^{-4}$.

\subsection{Integrated Radar and Communications}
    Due to the operating frequency bands of radar and communication system might be overlapped, particularly in the millimeter-wave spectrum, Sodagari~\cite{3} suggest a coexistent system of radar and communications by spectrum sharing technology. However, sharing spectrum inevitably cause mutual interference between radar and communications. Li, $et$ $al$.~\cite{52} employ the MC approach to eliminate interference between a special class of colocated MIMO radar and MIMO communication system. Moreover, it can improve transmission efficiency when the receive antennas communicate with the fusion center via only sending a small number of samples to fusion center. Sun~\cite{53} explain when the number of targets is less than the number of transmit and receive antennas, the data matrix at receiver possesses the low-rank and strong incoherence properties.
     \begin{figure}[!htbp]
     \begin{minipage}[b]{0.5\linewidth}
      {\includegraphics[scale=0.35]{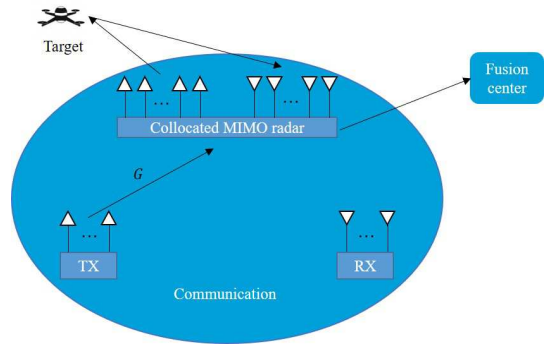}}
     \end{minipage}
     \caption{Colocated MIMO radar system sharing spectrum with MIMO communication system.}
     \end{figure}

    Fig. 7 depicts the coexistence of colocated MIMO radar and MIMO communication system. Herein, we use $\bm G\in \mathbb {R}^{M_r\times L}$ to denote the interference from the TX antennas of communication, and $\bm Y_R\in \mathbb {R}^{M_r\times L} = \bm X +\bm G$ to be the receive signal at the radar receiver where $\bm X$ is the original (unpolluted) signal and its rank is $K$ being the number of the targets, $M_r$ and $L$ are the numbers of receive antennas and the number of samples, respectively. After sub-sampling the signals impinging upon the radar receiver antennas, the sparse data $\bm H\odot\bm Y_R$ will be delivered to the fusion center. The sub-sampling rate is defined as $\left |\Omega \right |/|M_r\times L|$. At the fusion center, the receive signal is $\bm Y = \bm H \odot(\bm X + \bm G)+\bm N$ where $\bm N$ is the noise acquired during transmission, and then the original signal $\bm X$ can be recovered via~(\ref{E25}).

    Fig. 8 plots the normalized RMSE versus interference. As shown in Fig. 8, we can know that the MC technique is able to effectively suppress interference. Regarding the effect of sub-sampling ratio on the normalized RMSE, it is shown in Fig.~9. To compromise between normalized RMSE and sub-sampling ratio, the 50$\%$ sub-sampling ratio is a good choice.
     \begin{figure}[!htbp]
     \centering
     \begin{minipage}[b]{0.5\linewidth}
      {\includegraphics[scale=0.5]{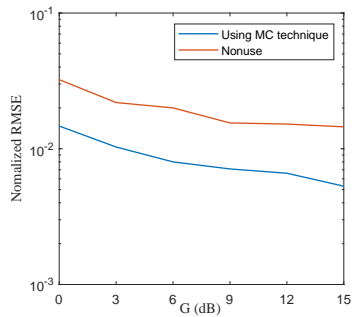}}
     \end{minipage}
     \caption{Normalized RMSE versus interference $\bm G$. $M_r=40$, $L = 128$, $K=2$, $\bm N$ being the 6dB Gaussian noise, and sub-sampling ratio being 50$\%$.}
     \end{figure}

     \begin{figure}[!htbp]
     \centering
     \begin{minipage}[b]{0.5\linewidth}
      {\includegraphics[scale=0.5]{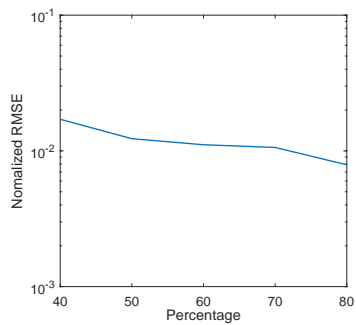}}
     \end{minipage}
     \caption{Normalized RMSE versus sub-sampling percentages. $M_r=40$, $L = 128$, $K=2$, $\bm G$ and $\bm N$ being the 0dB and 6dB Gaussian noises, respectively.}
     \end{figure}
\subsection{Traffic Sensing}
    Vehicular $ad$ $hoc$ network (VANET) is an efficient and economical technology compared to static detectors including cameras and underground inductive loops. It employs probe vehicles (PVs) (e.g., cars and buses) to sense and upload the traffic conditions. However, only fragments of real-time traffic information can be collected because PVs are unevenly distributed in the city. So, it is not efficient and accurate to estimate the traffic information. To cope with this challenge, Du $et$ $al$.~\cite{2} proposed to utilize the MC to determine real-time traffic conditions by using 20\% of the traffic data.

    Usually, vehicles equipped with onboard unit act as mobile sensors to collect the traversal time of roads. There are two phases in sensing process, including local sensing and data aggregation~\cite{2}. In the first stage, PVs record traffic information periodically, defined as ${\rm Traf}(j,n,t)$ that means traversal time $t$ of the $j$th link of the $n$th PV where link is defined as the road connecting two adjacent intersections. In the data aggregation stage, the sink nodes sample data from PVs over a $T$ period and compute the average traversal time (ATT) of each link. The ATT of the $i$th road during the $j$th sample period is defined as $x_{i,j}=\frac {1}{|N_i|}\sum_{n\in N_i,jT\leq t_j\leq (j+1)T}{\rm Traf}(i,n,t_j)$, where $\bm N_i$ is the set of vehicles that reports the information of the $i$th link and $|N_i|$ means the number of elements in the set. The ATT's $x_{i,j}$ consist of the sparse traffic condition matrix (TCM), of which row and column represent the link ID and sample time, respectively. It was revealed in~\cite{49}  that the TCM satisfies the approximately low-rank property for a large amount of data. Furthermore, the parallel algorithm~\cite{50} or distributed algorithm~\cite{51} can be utilized to improve the computation efficiency for the large size of traffic data.
\subsection{Potential applications}
    Three types of applications have been introduced. In this section, we will describe another two potential applications. The first one is state estimation in power system. Accurate state estimation can help rationalize the distribution electricity in order to achieve energy savings and lower carbon footprints. While traditional state estimation methods require full network observability. However, it is difficult to obtain full observability due to limited sensors in the whole network. Meanwhile the estimated state also affected by incorrect data that will mislead managers. Based on the application in traffic sensing, MC method is expected to be utilized to estimate state in power system.

    The second one is human motion recovery. Human motion analysis has been used to study human behavior and drive the machine, specifically, medical rehabilitation, behavior analysis and man-machine interfaces. For instance, action analysis can help athletes correct their actions such that it is able to improve their performance. However, human motion capture is a complicated process. what is more, the professional instrument cannot capture accurate and complete motion data because of occlusion issues caused by human body or clothing. Therefore, human motion recovery is receiving increasing attention. Essentially, the human motion data are similar to image data, so MC technique has the potential to be an important tool to solve this challenge.
\section{Conclusions and future directions}
    This survey has provided a comprehensive review of the MC technique from the signal processing perspective, including the principles of MC approaches as its variants,
    representative algorithms and potential applications. Firstly, we have re-formulated the MC problem so that the model can be adopted in areas of signal processing and wireless communications. Secondly, the principles of the MC philosophies have been revisited with insights, including semidefinite programming, nuclear norm relaxation, robust PCA, matrix factorization, minimum rank approximation, $\ell_p$-norm minimization and adaptive outlier pursuing. Meanwhile, we have discussed their pros and cons, and their application situations, varying from noiseless, Gaussian noise to Gaussian mixture noise. Particularly, the mathematical interpretation is provided to address why $\ell_p$-norm is able to resist impulsive noise. Thirdly, we have summarized five state-of-the-art optimization algorithms which are grouped into gradient and non-gradient types. Fourthly, simulation results demonstrated the empirical performance of five different MC formulations excluding SDP due to limitations of its application. Ultimately. we have showcased three representative application, namely SAR imaging, traffic sensing and integrated radar and communications. At the same time, two potential application are also described. In practice, experiment results based on real-world data have shown that the MC technique is able to compresses data and suppress noise efficiently in communications field.

    The MC problem has been extensively studied for decades. There is an assumption to solve the MC problem in most of state-of-the-art MC algorithms. That is that we have known the rank of the matrix before calculation. Regarding $\ell_p$-norm minimization, although $\ell_1$-norm has a best performance on impulsive noise, it is not the first choice in Gaussian noise case. In practice, it is difficult to obtain information about the exact rank and the type of noise. An open question is that is it possible to automatically adjust the rank and $p$ parameters? Xu and Sun~\cite{59} proposed a model-driven deep-learning framework which can learn the undetermined parameters autonomously during calculating target parameters. This framework may be a beacon for researchers to address these two challenges. We hope this tutorial article will serve as a good point for readers who would like to study the MC problem or apply the MC technique to their applications.

\section*{Acknowledgment}
The work described in this paper was supported by the National Natural Science Foundation of China under Grants U1713217 and U1501253.


\ifCLASSOPTIONcaptionsoff
  \newpage
\fi



%

%

\begin{IEEEbiography}[{\includegraphics[width=1in,height=1.25in,clip,keepaspectratio]{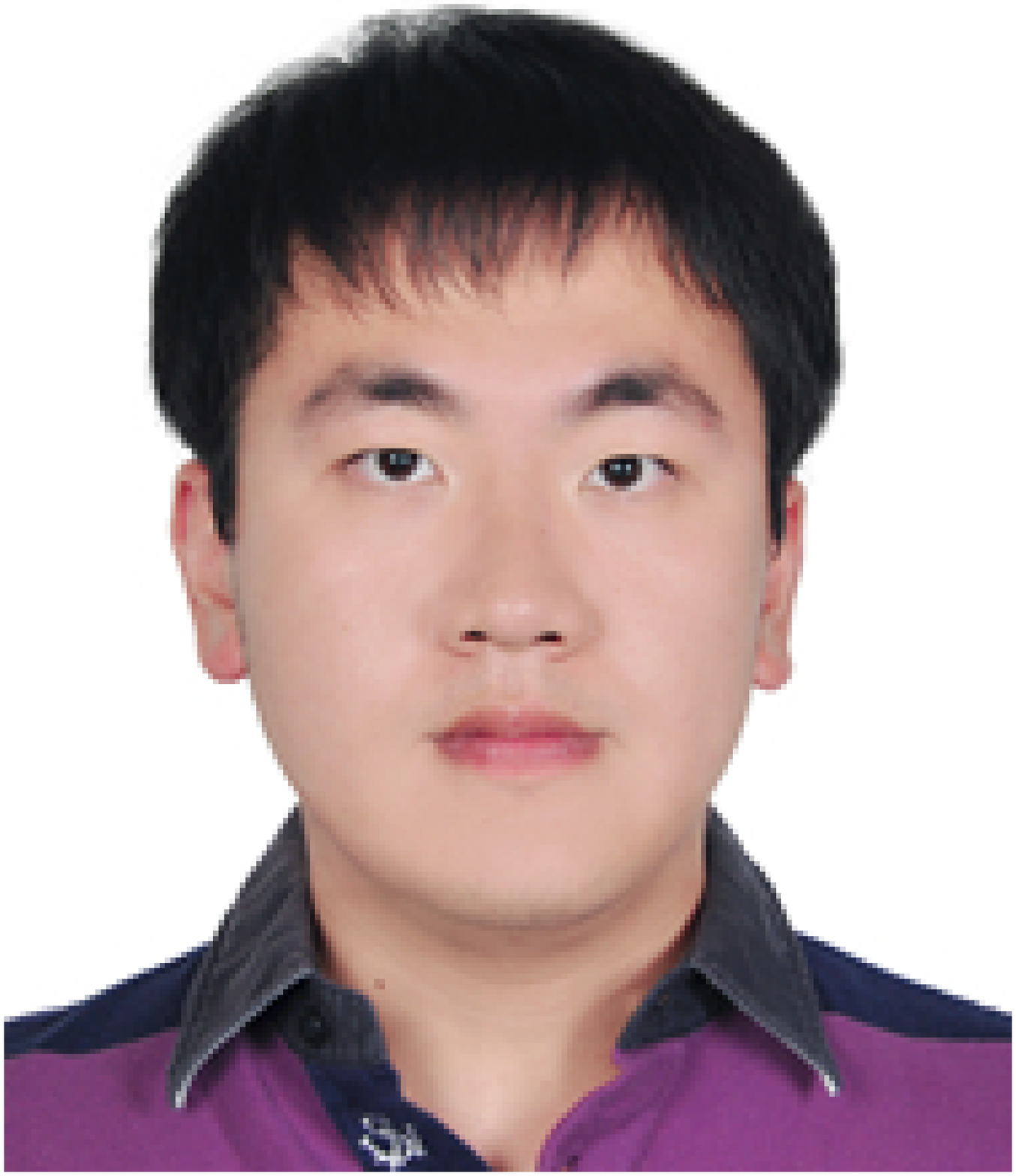}}]{Xiao~Peng~Li}
was born in Hebei, China in 1991. He received the B.E. degree in Electronic Science and Technology from Yanshan University, Qinhuangdao, China, and M.Sc. degree in Electronic Information Engineering from City University of Hong Kong, Hong Kong in 2015 and 2018, respectively. He is currently a Research Assistant with the College of Information Engineering, Shenzhen University, China.

His research interests are signal processing, sparse matrix completion and their applications in wireless communications.
\end{IEEEbiography}

\begin{IEEEbiography}[{\includegraphics[width=1in,height=1.25in,clip,keepaspectratio]{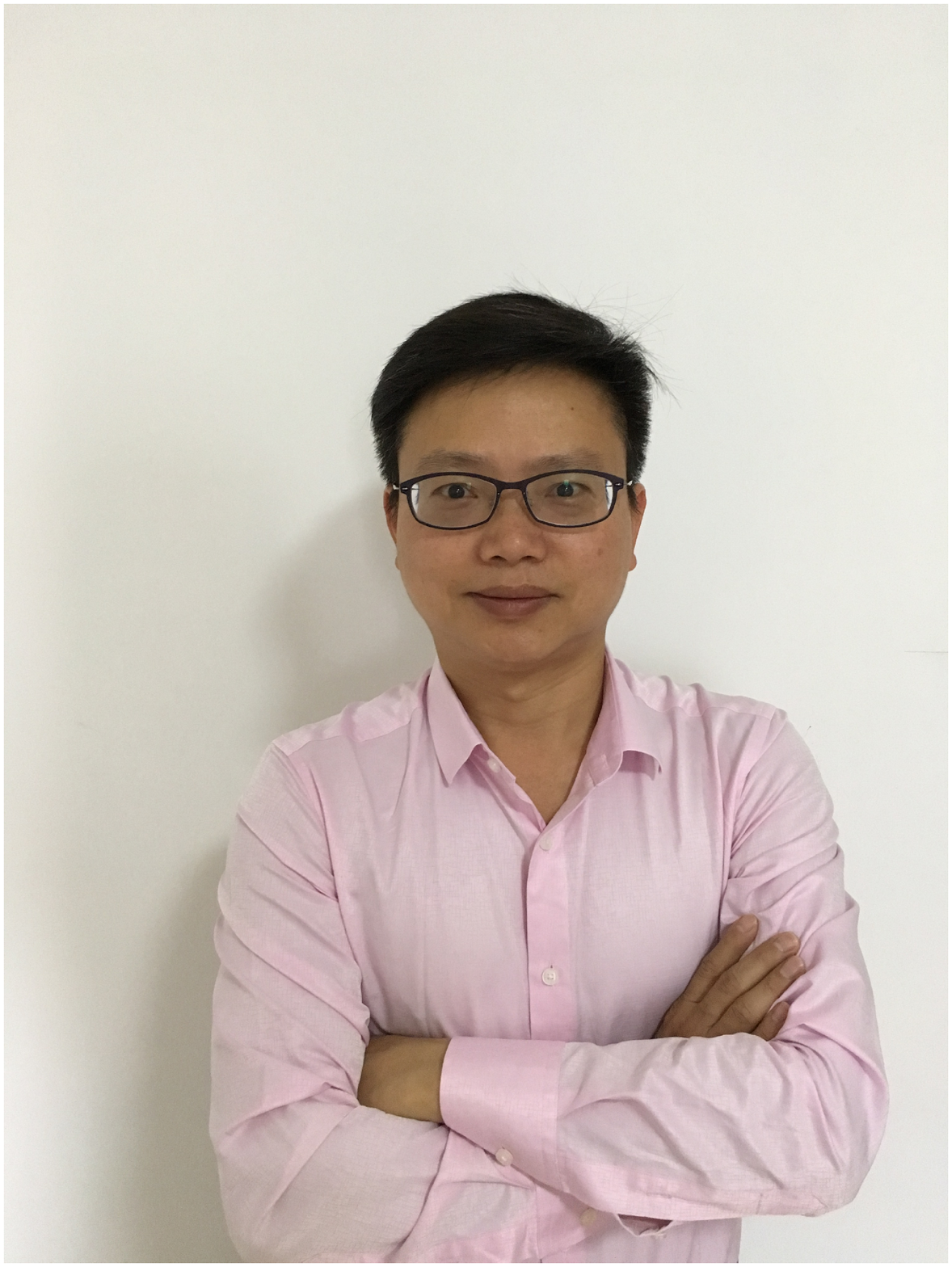}}]{Lei Huang}
(M'07-SM'14) was born in Guangdong, China. He received the B.Sc., M.Sc., and Ph.D. degrees in electronic engineering from Xidian University, Xi’an, China, in 2000, 2003, and 2005, respectively.

From 2005 to 2006, he was a Research Associate with the Department of Electrical and Computer Engineering, Duke University, Durham, NC, USA. From 2009 to 2010, he was a Research Fellow with the Department of Electronic Engineering, City University of Hong Kong, and a Research Associate with the Department of Electronic Engineering, Chinese University of Hong Kong. From 2012 to 2014, he was a Professor with the Department of Electronic and Information Engineering, Harbin Institute of Technology Shenzhen Graduate School. Since 2014, he has been with the College of Information Engineering, Shenzhen University, where he is currently a Distinguished Professor. His research interests include spectral estimation, array signal processing, statistical signal processing, and their applications in radar, and navigation and wireless communications.

He has been on the editorial boards of the IEEE Transactions on Signal Processing (2015-present), Elsevier-Digital Signal Processing (2012-present) and IET Signal Processing (2017-present). He has been an elected member of the Sensor Array and Multichannel (SAM) Technical Committee of the IEEE Signal Processing Society (2016-present). He was elected an IET Fellow in 2018.
\end{IEEEbiography}
\begin{IEEEbiography}[{\includegraphics[width=1in,height=1.25in,clip,keepaspectratio]{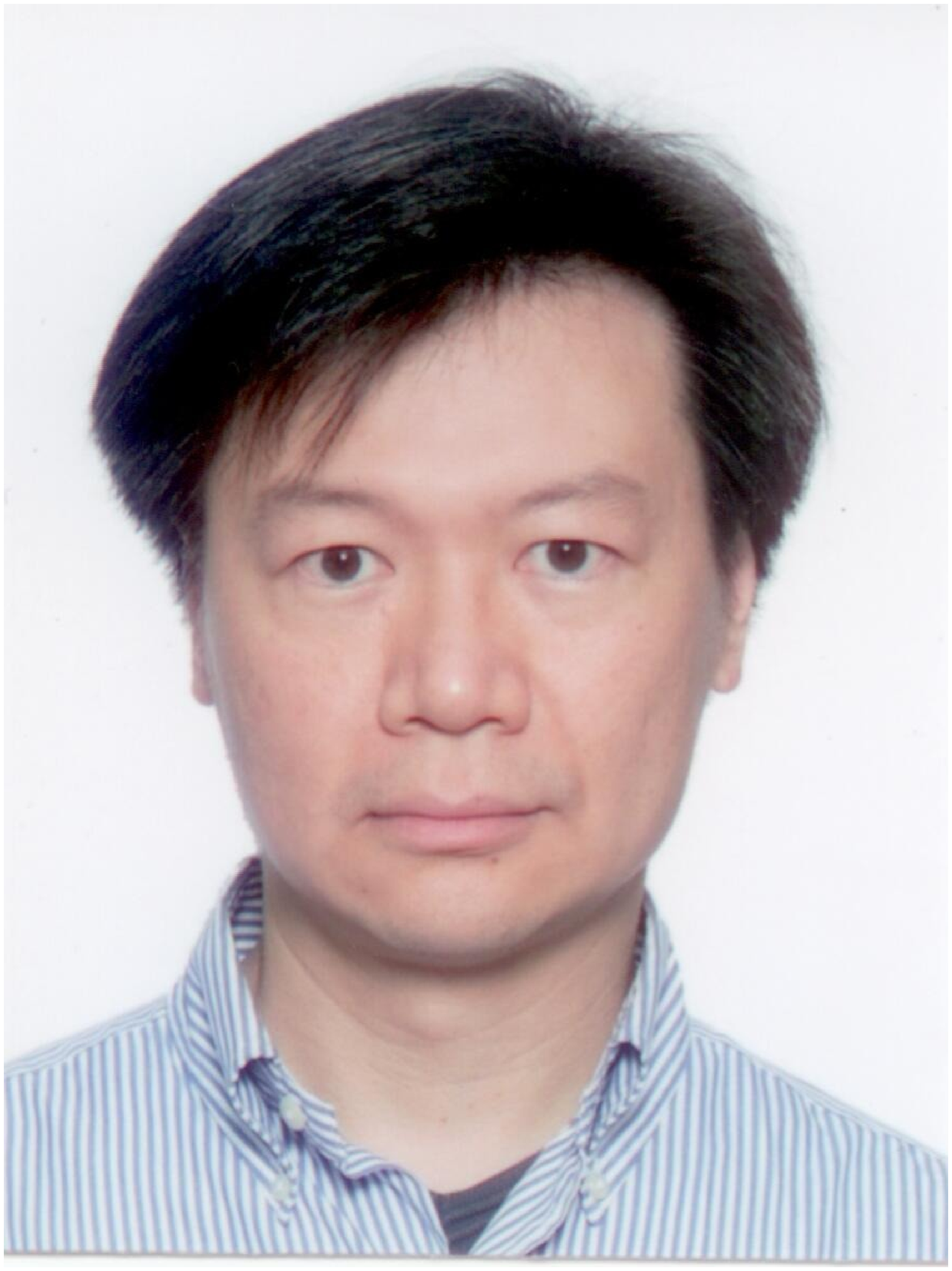}}]{Hing Cheung So}
(S'90-M'95-SM'07-F'15) was born in Hong Kong. He received the B.Eng. degree from the City University of Hong Kong and the Ph.D. degree from The Chinese University of Hong Kong, both in electronic engineering, in 1990 and 1995, respectively. From 1990 to 1991, he was an Electronic Engineer with the Research and Development Division, Everex Systems Engineering Ltd., Hong Kong. During 1995-1996, he was a Postdoctoral Fellow with The Chinese University of Hong Kong. From 1996 to 1999, he was a Research Assistant Professor with the Department of Electronic Engineering, City University of Hong Kong, where he is currently a Professor. His research interests include detection and estimation, fast and adaptive algorithms, multidimensional harmonic retrieval, robust signal processing, source localization, and sparse approximation.

He has been on the editorial boards of $IEEE$ $Signal$ $Processing$ $Magazine$ (2014-2017), $IEEE$ $Transactions$ $on$ $Signal$ $Processing$ (2010-2014), $Signal$ $Processing$ (2010-), and $Digital$ $Signal$ $Processing$ (2011-). He was also Lead Guest Editor for $IEEE$ $Journal$ $of$ $Selected$ $Topics$ $in$ $Signal$ $Processing$, special issue on “Advances in Time/Frequency Modulated Array Signal Processing” in 2017. In addition, he was an elected member in Signal Processing Theory and Methods Technical Committee (2011-2016) of the IEEE Signal Processing Society where he was chair in the awards subcommittee (2015-2016).
\end{IEEEbiography}

\begin{IEEEbiography}[{\includegraphics[width=1in,height=1.25in,clip,keepaspectratio]{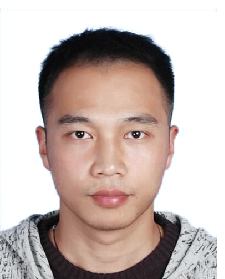}}]{Bo Zhao}
(M'15) was born in Henan, China, in 1986. He received the B.Sc. and Ph.D. degrees from Xidian University, Xi'an, China, in 2010 and 2015, respectively.
From 2015 to 2018, he was a postdoctoral researcher with the College of Information Engineering, Shenzhen University. He is currently an Assistant Professor with the College of Information Engineering, Shenzhen University.

His research interests include radar imaging, SAR countermeasure, and compressive sensing.
\end{IEEEbiography}
\end{document}